\documentclass[aps,prl,reprint,superscriptaddress,amsmath,amssymb,floatfix]{revtex4-2}

\usepackage{graphicx}
\usepackage{booktabs}
\usepackage{multirow}
\usepackage{dcolumn}
\usepackage{makecell}
\usepackage{mathtools}
\usepackage{tabularx}
\usepackage{colortbl}
\usepackage{physics}

\usepackage[colorlinks,linkcolor=blue,anchorcolor=blue,citecolor=blue, urlcolor=blue]{hyperref}

\usepackage{multirow}

\begin{document}

\title{Inverse-k Primordial Oscillations from a Symbolic Regression Search}

\author{Ze-Yu Peng}
\email{pengzeyu23@mails.ucas.ac.cn}
\affiliation{School of Physical Sciences, University of Chinese Academy of Sciences, Beijing 100049, China}
\affiliation{International Centre for Theoretical Physics Asia-Pacific, University of Chinese Academy of Sciences, 100190 Beijing, China}

\author{Qing-Yu Lan}
\email{lanqingyu19@mails.ucas.ac.cn} \affiliation{School of
    Physical Sciences, University of Chinese Academy of Sciences,
Beijing 100049, China}

\author{Yun-Song Piao}
\email{yspiao@ucas.ac.cn}
\affiliation{School of Physical Sciences, University of Chinese Academy of Sciences, Beijing 100049, China}
\affiliation{International Centre for Theoretical Physics Asia-Pacific, University of Chinese Academy of Sciences, 100190 Beijing, China}
\affiliation{School of Fundamental Physics and Mathematical Sciences, Hangzhou Institute for Advanced Study, UCAS, Hangzhou 310024, China}
\affiliation{Institute of Theoretical Physics, Chinese Academy of Sciences, P.O. Box 2735, Beijing 100190, China}

\begin{abstract}
    Oscillatory features in the primordial power spectrum,
    potential signatures of new physics in the early universe,
    are usually searched for using fixed templates. In this
    work, we perform a template-free search for primordial features
    using symbolic regression. We find that both Planck and the combined
    Planck+ACT+SPT-3G datasets independently select an
    inverse-$k$ oscillation, $\cos(B/k)$ with $B\simeq4\,\mathrm{Mpc}^{-1}$,
    as the leading low-complexity feature. Comparing this inverse-$k$ template
    with standard linear and logarithmic oscillating templates, we find that it
    fits the data best, showing a weak preference for a non-zero amplitude.
    Our results show that symbolic regression as a powerful machine learning technique can provide an interpretable, model-independent approach to cosmological discovery.
\end{abstract}

\maketitle

\emph{Introduction.---} Inflation
\cite{Guth:1980zm,Linde:1981mu,Albrecht:1982wi,Starobinsky:1980te}, the
standard paradigm of the very early universe, predicts a nearly
scale-invariant primordial power spectrum (PPS) of scalar perturbations,
which is in good agreement with the Planck cosmic microwave background
(CMB) observations \cite{Planck:2018vyg}. Nevertheless, departures from
the slow-roll dynamics can imprint oscillatory features on the PPS
\cite{Starobinsky:1992ts,Adams:2001vc,Wang:2002hf,Gong:2005jr,Slosar:2019gvt,Achucarro:2022qrl}, while remaining consistent with current observations.
The search for such features offers a unique window into the new physics in the primordial universe \cite{Chen:2012ja,Chen:2018cgg}.

Previous searches for primordial oscillatory features have mostly used
linear and logarithmic oscillation templates, and have found no evidence for a departure from the power-law spectrum
\cite{Easther:2004vq,Meerburg:2013cla,Meerburg:2013dla,Peiris:2013opa,Easther:2013kla,Aich:2011qv,Planck:2018jri,Beutler:2019ojk,Ballardini:2022wzu,Mergulhao:2023ukp}.
The tightest constraint to date comes from the combination of Planck
with the latest ACT \cite{ACT:2025blo,ACT:2025nti} and SPT-3G \cite{SPT-3G:2025bzu} data \cite{Peng:2025vda,Nerval:2026iev}.
See also \cite{Braglia:2021ckn,Braglia:2021sun,Braglia:2021rej,Braglia:2022ftm,Petretti:2024mjy} for constraints on other realistic models of primordial features.

Symbolic regression (SR) \cite{schmidt2009distilling,cranmer2023}
has emerged as a powerful tool for scientific
discovery. By searching for explicit analytical expressions that
optimize a given objective, SR enables the discovery of
interpretable models directly from data without relying on
predefined functional forms or black-box approximators.
SR has been shown to rediscover known physical laws directly from experimental data \cite{Udrescu:2019mnk}.
It has recently been applied to a variety of problems in cosmology, e.g.~\cite{Sousa-Neto:2025gpj,Koksbang:2026wvh} for the reconstruction of the expansion history and \cite{Bartlett:2022kyi} for an exhaustive SR approach.

In this work, we perform a template-free search for primordial features using SR. We consider Planck and its combination with the latest ACT and SPT-3G data, and find that both datasets independently select an inverse-$k$ oscillation, $\cos(B/k)$ with $B\simeq4\,\mathrm{Mpc}^{-1}$, as the leading low-complexity feature. We further compare this inverse-$k$ template with the standard linear and logarithmic templates, and find that it fits the data better and shows a weak preference for a nonzero amplitude, while the standard templates remain consistent with zero.


\emph{Methods and datasets.---} We write the primordial power spectrum (PPS) as
\begin{equation}
    P_{\mathcal R}(k) =
    P_{\mathcal R,0}(k)\left[1+f(k)\right],
    \quad
    P_{\mathcal R,0}(k)=A_s\left(\frac{k}{k_*}\right)^{n_s-1},
\end{equation}
with $k_*=0.05\,{\mathrm{Mpc}^{-1}}$. Here $f(k)$ is the primordial feature we search
for, and $f(k)=0$ recovers the standard power-law spectrum $P_{\mathcal R,0}$.

We search for analytic expressions for $f(k)$ using symbolic regression
as implemented in \texttt{PySR}~\cite{cranmer2023}, with the CMB $\chi^2$ as the
loss function. The CMB spectra are computed using CLASS
\cite{Blas:2011rf}, and all the cosmological and nuisance parameters are
fixed to their best-fit values for the standard power-law spectrum. See Supplemental Material~\cite{SupplementalMaterial} for details of the SR search.

We consider two CMB datasets, Planck and SPA (Planck$+$SPT$+$ACT). The
Planck dataset includes the full high-$\ell$ Plik-lite TTTEEE and low-$\ell$
Commander TT likelihoods \cite{Planck:2018vyg}. The SPA dataset combines the SPT-3G D1
\cite{SPT-3G:2025bzu} and ACT DR6 \cite{ACT:2025blo,ACT:2025nti} data
with Planck. Following \cite{ACT:2025blo, SPT-3G:2025bzu}, the Planck
high-$\ell$ data in SPA are cropped to $\ell_{\mathrm{max}}=1000$ for TT and $600$
for TE and EE; we further restrict SPT and ACT to $\ell\le2500$, matching
the scales of Planck and avoiding the modeling
uncertainties at smaller scales~\cite{Stahl:2025qru,Calderon:2025xod}.

\emph{SR results.---}

\begin{figure}[htbp]
    \centering
    \includegraphics[width=\linewidth]{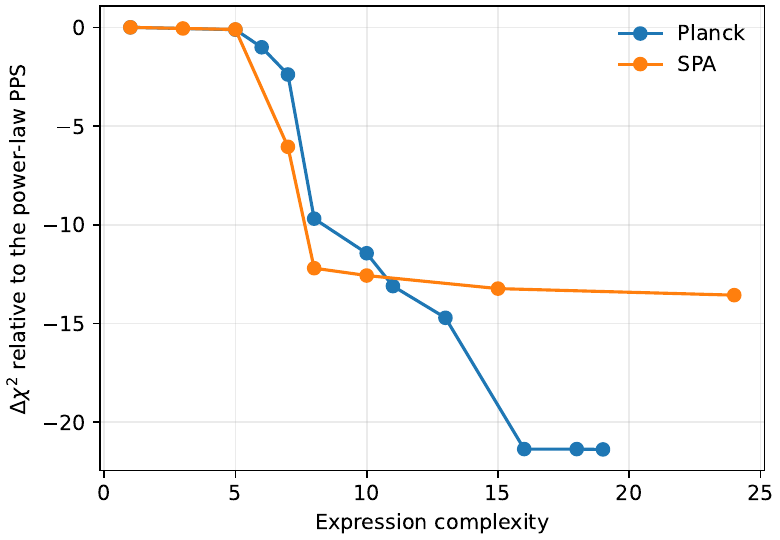}
    \caption{
        Improvement of the CMB likelihood along the SR
        Pareto front relative to the standard power-law spectrum.
    }
    \label{fig:pysr_pareto}
\end{figure}

\begin{table}[htbp]
    \centering
    \begin{tabular}{|c|c|l|}
        \hline
        Complexity & $\Delta\chi^2_{\mathrm{Planck}}$ & $f(k)$ \\
        \hline
        8  & $-9.69$  & $-0.0299\cos(4.101/k)$ \\
        10 & $-11.44$ & $-0.0339\cos(4.150/k-0.859)$ \\
        11 & $-13.10$ & $k\sin(-0.868/k^2)$ \\
        16 & $-21.35$ & $0.0924\sin[(2.990+0.5248/k)/k]$ \\
        \hline
    \end{tabular}
    \caption{
        Pareto-front expressions from the Planck symbolic-regression run.
        The improvement $\Delta\chi^2$ is measured relative to the no-feature power-law spectrum.
    }
    \label{tab:pysr_planck}
\end{table}

\begin{table}[htbp]
    \centering
    \begin{tabular}{|c|c|l|}
        \hline
        Complexity & $\Delta\chi^2_{\mathrm{SPA}}$ & $f(k)$ \\
        \hline
        7  & $-6.05$  & $-0.0106\sin(-390.0 k)$ \\
        8  & $-12.20$ & $-0.0209\sin(4.231/k)$ \\
        10 & $-12.57$ & $0.0209\sin(4.178/k-2.594)$ \\
        15 & $-13.23$ & $0.0218\sin(4.173/k+4.173/e^k)$ \\
        \hline
    \end{tabular}
    \caption{
        Pareto-front expressions from the SPA symbolic-regression run.
        The improvement $\Delta\chi^2$ is measured relative to the no-feature power-law spectrum.
    }
    \label{tab:pysr_spa}
\end{table}

In Fig.~\ref{fig:pysr_pareto}, we show the Pareto fronts obtained from
the SR search for the Planck and SPA datasets, with some representative
expressions listed in Tables~\ref{tab:pysr_planck} and
\ref{tab:pysr_spa}; the complete results are given in
Supplemental Material~\cite{SupplementalMaterial}. For both datasets, the likelihood
improvement slows down around complexity~8 (C8), where the two runs converge
to the same form, an inverse-$k$ oscillation with similar frequency,
\begin{align}
    f_{\mathrm{Planck}}^{(8)}(k)
    &\simeq -0.0299\,\cos\left(\frac{4.101}{k}\right),\\
    f_{\mathrm{SPA}}^{(8)}(k)
    &\simeq -0.0209\,\sin\left(\frac{4.231}{k}\right),
\end{align}
with $k$ measured in ${\mathrm{Mpc}^{-1}}$.

Beyond C8, the two
datasets behave differently. For SPA, the front saturates: the
higher-complexity C10, C15 and C24 expressions are essentially the same
inverse-$k$ oscillation, differing at most by a small $e^{-k}$ term in
the phase. For Planck, the likelihood keeps improving, but the
additional gain comes from inverse-power terms such as $1/k^2$, which
oscillate increasingly rapidly toward small $k$ and fall below the
sampling resolution of our CLASS runs; we therefore treat these
high-complexity Planck expressions as numerically unreliable, see details in Supplemental Material~\cite{SupplementalMaterial}.

\begin{figure*}[htbp]
    \centering
    \includegraphics[width=\linewidth]{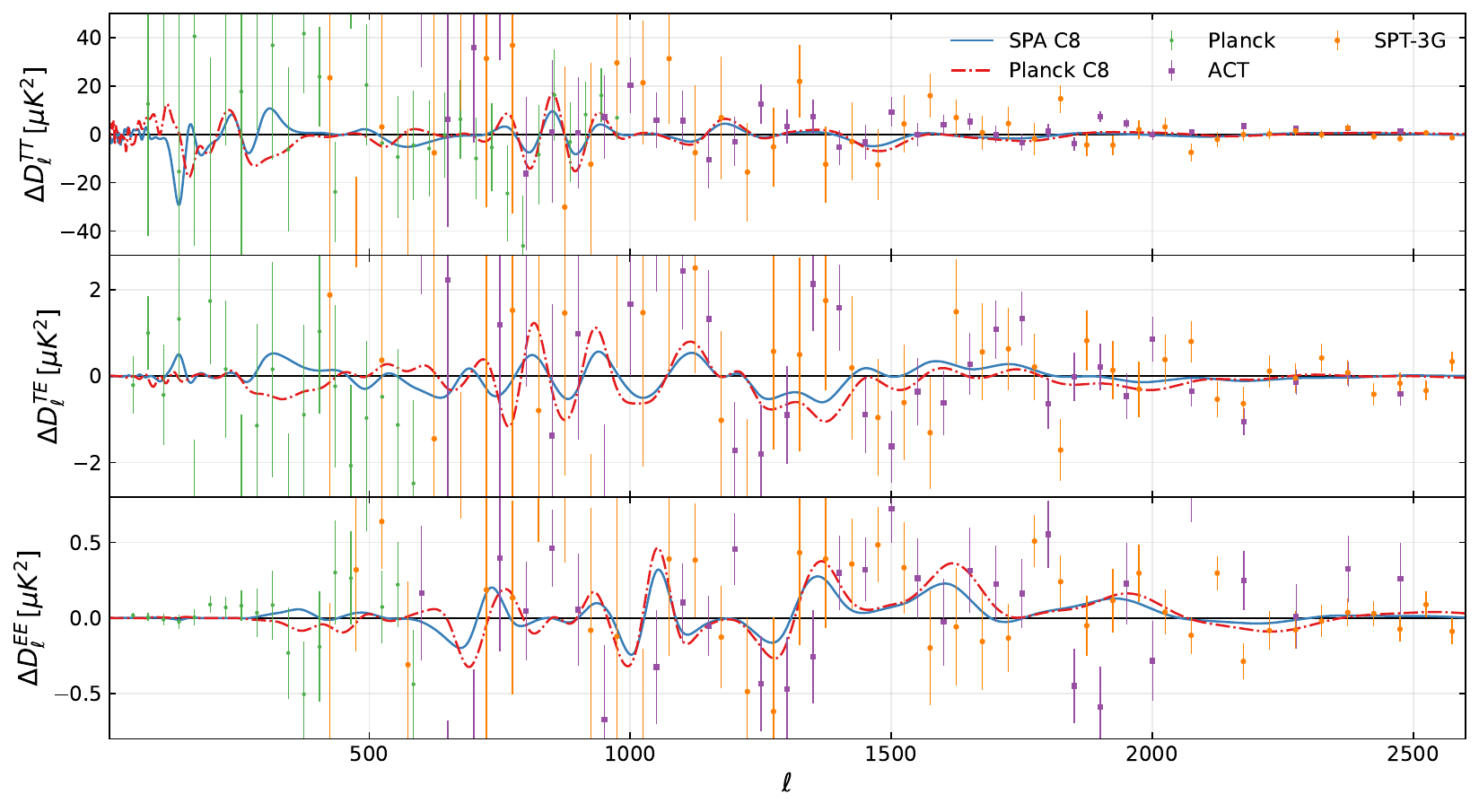}
    \caption{
        CMB residuals for SR expressions with complexity $8$, relative to the
        SPA power-law baseline. We also show the Planck (cropped), ACT and SPT-3G data bins.
    }
    \label{fig:cmb_sr_residuals}
\end{figure*}

In Fig.~\ref{fig:cmb_sr_residuals}, we show the CMB residuals of the C8
expressions, projected from the PPS to the TT, TE and EE spectra
relative to the SPA power-law baseline. The Planck and SPA expressions
yield a similar oscillatory pattern over the range
$800\lesssim\ell\lesssim2000$, which tracks the oscillations in the
measured residuals from SPT and ACT, most clearly in the TE and EE
polarization spectra.

\emph{Comparison with standard templates.---} The SR search suggests an inverse-$k$ oscillation in the PPS.
We compare it with the commonly used linear and logarithmic oscillation
templates.

\begin{table}[htbp]
    \centering
    \begin{tabular}{|c|c|}
        \hline
        Parameter & Prior \\
        \hline
        $A_X$ & $[0,0.5]$ \\
        $\omega_X$ & $[1,100]$ \\
        $\phi_X$ & $[0,2\pi]$ \\
        \hline
    \end{tabular}
    \caption{Priors for the three templates}\label{tab:priors}
\end{table}

The three templates of primordial features are defined as
\begin{equation}
    f_X(k) =
    A_X\cos\left[\omega_X g_X(k)+\phi_X\right],
\end{equation}
with
\begin{equation}
    g_{\mathrm{lin}}(k)=\frac{k}{k_*},\quad
    g_{\mathrm{log}}(k)=\log\frac{k}{k_*},\quad
    g_{\mathrm{inv}}(k)=\frac{k_*}{k}.
\end{equation}
We use the same flat priors for all three templates, listed in
Table~\ref{tab:priors}\footnote{Note that the MCMC results depend on the choice of priors, in particular on $\omega_X$, which is not equivalent across the three templates; the comparison should therefore be regarded as heuristic.}.
For each template, we run a separate Markov
chain Monte Carlo (MCMC) analysis over $(A_X,\omega_X,\phi_X)$ on the SPA
dataset using \texttt{Cobaya} \cite{Torrado:2020dgo}, with the background
and nuisance parameters fixed as in the SR search. It has been shown that the variation of the background and nuisance parameters has a negligible effect on the feature constraints \cite{Peng:2025vda}.

\begin{figure}[htbp]
    \centering
    \includegraphics[width=\linewidth]{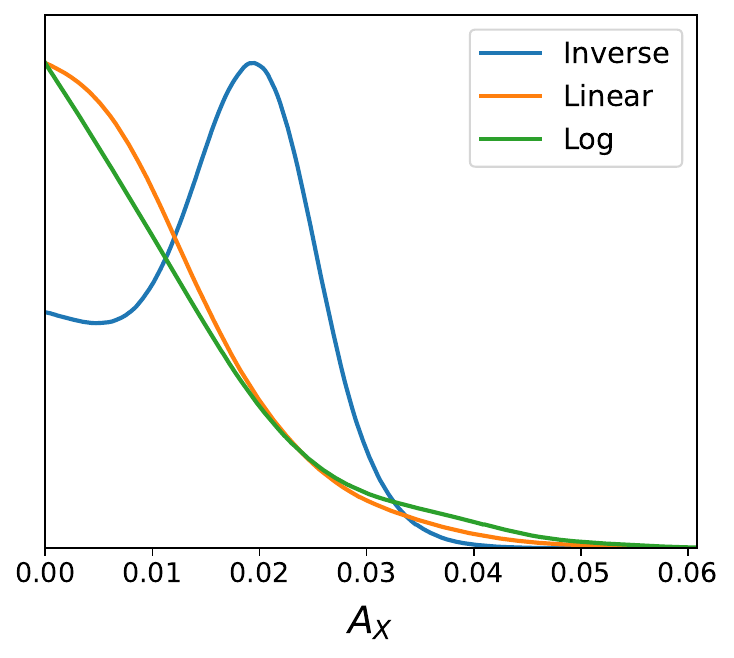}
    \caption{
        Marginalized posteriors of the oscillation amplitude $A_X$ for
        the linear, logarithmic and inverse-$k$ templates on the SPA
        dataset.
    }
    \label{fig:A_X}
\end{figure}

Figure~\ref{fig:A_X} shows the marginalized posteriors of the amplitude
$A_X$.
We find a $1\sigma$ lower bound for the amplitude of the inverse-$k$ template, $A_X = 0.0163^{+0.0099}_{-0.0071}$, showing a weak preference for a nonzero amplitude,
while the linear and logarithmic templates are consistent with zero and give only upper limits.

\begin{table*}[htbp]
    \centering
    \begin{tabular}{|c|c|c|c|c|c|c|c|}
        \hline
        Template & $A_X$ & $\omega_X$ & $\phi_X$ & $\Delta\chi^2$ & Planck & SPT-3G & ACT \\
        \hline
        baseline & 0 & -- & -- & 0.00 & 243.76 & 109.90 & 110.29 \\
        linear & 0.0113 & 18.69 & 6.00 & -7.69 & 239.26 & 108.64 & 108.35 \\
        log & 0.0309 & 83.14 & 1.64 & -5.85 & 241.51 & 110.40 & 106.18 \\
        inverse & 0.0203 & 83.32 & 2.24 & -12.66 & 240.78 & 103.94 & 106.55 \\
        \hline
    \end{tabular}
    \caption{
        Best-fit template parameters and component $\chi^2$ values for the
        SPA dataset.
    }
    \label{tab:template_chi2}
\end{table*}

\begin{figure*}[htbp]
    \centering
    \includegraphics[width=\linewidth]{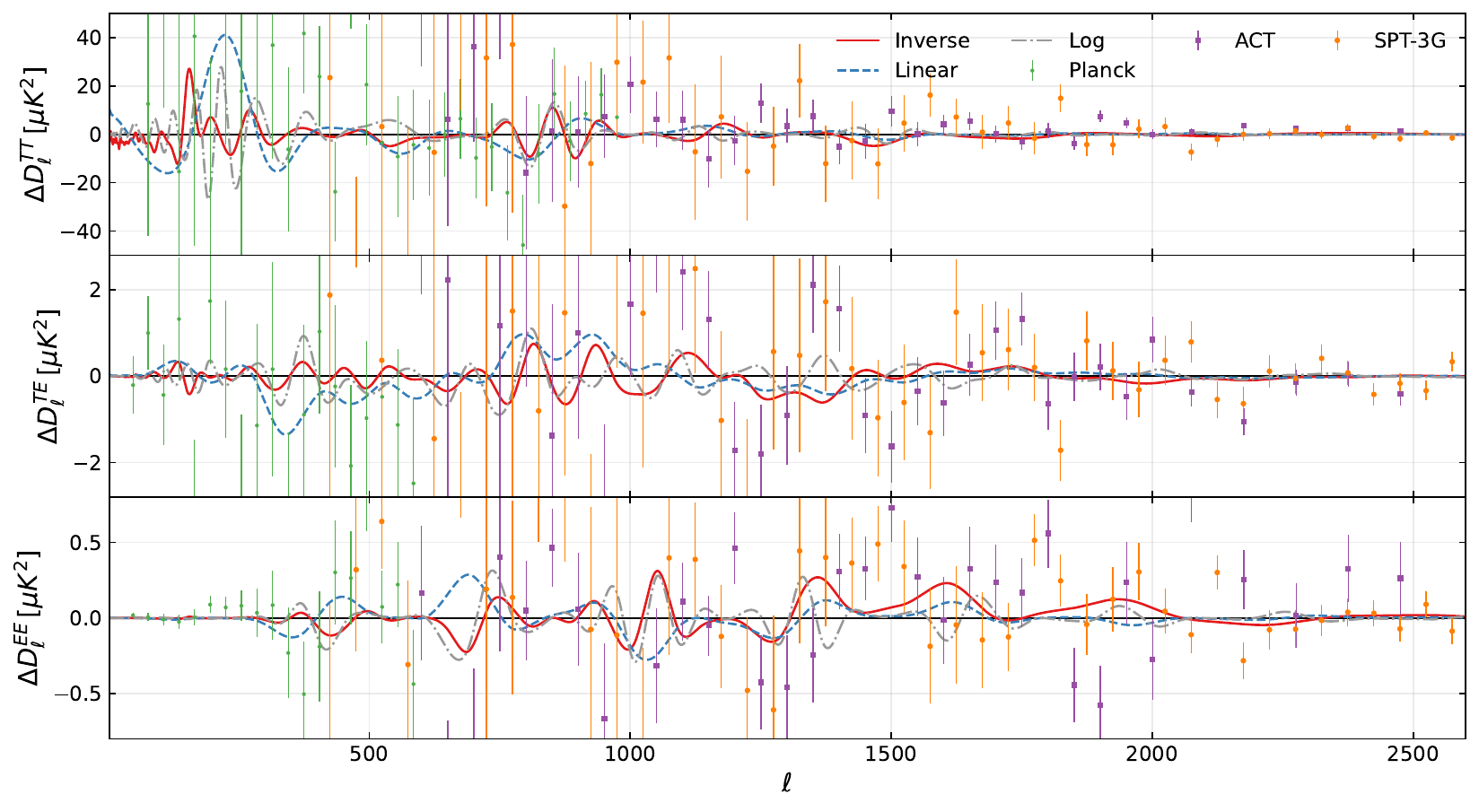}
    \caption{
        CMB residuals of the best-fit linear, logarithmic and inverse-$k$
        templates relative to the no-feature SPA baseline, compared with
        the Planck (cropped), ACT and SPT-3G data.
    }
    \label{fig:template_residuals}
\end{figure*}

The best-fit $\chi^2$ values are summarized in
Table~\ref{tab:template_chi2}. All three templates improve over the
no-feature baseline, but the inverse-$k$ template gives the largest
improvement, $\Delta\chi^2\simeq-12.6$, against $-7.69$ for the linear and
$-5.8$ for the logarithmic template. This improvement is dominated by
the small-scale SPT-3G and ACT data. The corresponding CMB residuals are
shown in Fig.~\ref{fig:template_residuals}.

\emph{Discussion.---} The search for primordial features offers a unique window into the new physics of the primordial universe.
In this work, we have performed a template-free search for primordial features using symbolic regression. For both Planck and its combination with the latest ACT and SPT-3G data, the search independently selects an inverse-$k$ oscillation as the leading low-complexity feature.
We further compare this inverse-$k$ template with the standard linear and logarithmic oscillation templates using MCMC. We find a weak preference for a nonzero amplitude for the inverse-$k$ oscillation, while the standard templates remain consistent with zero.

According to the primordial standard-clocks formalism \cite{Chen:2012ja,Chen:2018cgg}, a $k^{1/p}$ oscillation in the PPS can be generated by a massive field oscillating in a background with scale factor $a(t)\sim t^p$. Therefore, our inverse-$k$ oscillation may be interpreted as an early fast expansion phase with a slowly increasing Hubble parameter ($p=-1$, $t<0$ and $w<-1$), as may occur in NEC-violating or phantom-like early-universe scenarios \cite{Piao:2004tq,Baldi:2005gk, Cai:2020qpu,Ye:2023tpz,Cai:2023uhc,Pan:2024ydt}.
It should be noted that we regard this connection as heuristic, since our results only identify the oscillation phase rather than the full standard-clocks signal. Specific model implementation and observational tests are left for future work.

Our results demonstrate that symbolic regression provides a useful and complementary approach to the search for primordial features, able to identify forms beyond the standard templates.
It is expected that future CMB experiments, such as the Simons Observatory \cite{SimonsObservatory:2018koc} and CMB-S4 \cite{Abazajian:2019eic}, which provide improved measurements on small scales, will further refine our understanding of primordial features.

\begin{acknowledgments}
    This work is supported by NSFC, No.12475064, 12075246, National Key Research and Development Program of China, No. 2021YFC2203004, and the Fundamental Research Funds for the Central Universities.
    We acknowledge the use of high performance computing services provided by the International Centre for Theoretical Physics Asia-Pacific cluster.
\end{acknowledgments}

\bibliographystyle{apsrev4-1}
\bibliography{ref}

@article{Achucarro:2022qrl,
  archiveprefix = {arXiv},
  author        = {Ach{\'u}carro, Ana and others},
  eprint        = {2203.08128},
  month         = {3},
  primaryclass  = {astro-ph.CO},
  title         = {{Inflation: Theory and Observations}},
  year          = {2022}
}

@article{ACT:2025blo,
  archiveprefix = {arXiv},
  author        = {Louis, Thibaut and others},
  collaboration = {Atacama Cosmology Telescope},
  doi           = {10.1088/1475-7516/2025/11/062},
  eprint        = {2503.14452},
  journal       = {JCAP},
  pages         = {062},
  primaryclass  = {astro-ph.CO},
  reportnumber  = {FERMILAB-PUB-25-0071-PPD},
  title         = {{The Atacama Cosmology Telescope: DR6 power spectra, likelihoods and {\ensuremath{\Lambda}}CDM parameters}},
  volume        = {11},
  year          = {2025}
}

@article{ACT:2025nti,
  archiveprefix = {arXiv},
  author        = {Calabrese, Erminia and others},
  collaboration = {Atacama Cosmology Telescope},
  doi           = {10.1088/1475-7516/2025/11/063},
  eprint        = {2503.14454},
  journal       = {JCAP},
  pages         = {063},
  primaryclass  = {astro-ph.CO},
  reportnumber  = {FERMILAB-PUB-25-0157-PPD},
  title         = {{The Atacama Cosmology Telescope: DR6 constraints on extended cosmological models}},
  volume        = {11},
  year          = {2025}
}

@article{Adams:2001vc,
  archiveprefix = {arXiv},
  author        = {Adams, Jennifer A. and Cresswell, Bevan and Easther, Richard},
  doi           = {10.1103/PhysRevD.64.123514},
  eprint        = {astro-ph/0102236},
  journal       = {Phys. Rev. D},
  pages         = {123514},
  reportnumber  = {CU-TP-1005},
  title         = {{Inflationary perturbations from a potential with a step}},
  volume        = {64},
  year          = {2001}
}

@article{Aich:2011qv,
  archiveprefix = {arXiv},
  author        = {Aich, Moumita and Hazra, Dhiraj Kumar and Sriramkumar, L. and Souradeep, Tarun},
  doi           = {10.1103/PhysRevD.87.083526},
  eprint        = {1106.2798},
  journal       = {Phys. Rev. D},
  pages         = {083526},
  primaryclass  = {astro-ph.CO},
  title         = {{Oscillations in the inflaton potential: Complete numerical treatment and comparison with the recent and forthcoming CMB datasets}},
  volume        = {87},
  year          = {2013}
}

@article{Albrecht:1982wi,
  author       = {Albrecht, Andreas and Steinhardt, Paul J.},
  doi          = {10.1103/PhysRevLett.48.1220},
  editor       = {Fang, Li-Zhi and Ruffini, R.},
  journal      = {Phys. Rev. Lett.},
  pages        = {1220--1223},
  reportnumber = {UPR-0185T},
  title        = {{Cosmology for Grand Unified Theories with Radiatively Induced Symmetry Breaking}},
  volume       = {48},
  year         = {1982}
}

@article{Ballardini:2022wzu,
  archiveprefix = {arXiv},
  author        = {Ballardini, Mario and Finelli, Fabio and Marulli, Federico and Moscardini, Lauro and Veropalumbo, Alfonso},
  doi           = {10.1103/PhysRevD.107.043532},
  eprint        = {2202.08819},
  journal       = {Phys. Rev. D},
  number        = {4},
  pages         = {043532},
  primaryclass  = {astro-ph.CO},
  title         = {{New constraints on primordial features from the galaxy two-point correlation function}},
  volume        = {107},
  year          = {2023}
}

@article{Beutler:2019ojk,
  archiveprefix = {arXiv},
  author        = {Beutler, Florian and Biagetti, Matteo and Green, Daniel and Slosar, An{\v{z}}e and Wallisch, Benjamin},
  doi           = {10.1103/PhysRevResearch.1.033209},
  eprint        = {1906.08758},
  journal       = {Phys. Rev. Res.},
  number        = {3},
  pages         = {033209},
  primaryclass  = {astro-ph.CO},
  title         = {{Primordial Features from Linear to Nonlinear Scales}},
  volume        = {1},
  year          = {2019}
}

@article{Braglia:2021ckn,
  archiveprefix = {arXiv},
  author        = {Braglia, Matteo and Chen, Xingang and Hazra, Dhiraj Kumar},
  doi           = {10.1088/1475-7516/2021/06/005},
  eprint        = {2103.03025},
  journal       = {JCAP},
  pages         = {005},
  primaryclass  = {astro-ph.CO},
  title         = {{Comparing multi-field primordial feature models with the Planck data}},
  volume        = {06},
  year          = {2021}
}

@article{Braglia:2021rej,
  archiveprefix = {arXiv},
  author        = {Braglia, Matteo and Chen, Xingang and Hazra, Dhiraj Kumar},
  doi           = {10.1103/PhysRevD.105.103523},
  eprint        = {2108.10110},
  journal       = {Phys. Rev. D},
  number        = {10},
  pages         = {103523},
  primaryclass  = {astro-ph.CO},
  reportnumber  = {IFT-UAM/CSIC-21-129},
  title         = {{Primordial standard clock models and CMB residual anomalies}},
  volume        = {105},
  year          = {2022}
}

@article{Braglia:2021sun,
  archiveprefix = {arXiv},
  author        = {Braglia, Matteo and Chen, Xingang and Hazra, Dhiraj Kumar},
  doi           = {10.1140/epjc/s10052-022-10461-3},
  eprint        = {2106.07546},
  journal       = {Eur. Phys. J. C},
  number        = {5},
  pages         = {498},
  primaryclass  = {astro-ph.CO},
  reportnumber  = {IFT-UAM/CSIC-21-128},
  title         = {{Uncovering the history of cosmic inflation from anomalies in cosmic microwave background spectra}},
  volume        = {82},
  year          = {2022}
}

@article{Braglia:2022ftm,
  archiveprefix = {arXiv},
  author        = {Braglia, Matteo and Chen, Xingang and Hazra, Dhiraj Kumar and Pinol, Lucas},
  doi           = {10.1088/1475-7516/2023/03/014},
  eprint        = {2210.07028},
  journal       = {JCAP},
  pages         = {014},
  primaryclass  = {astro-ph.CO},
  reportnumber  = {IFT-UAM/CSIC-22-125},
  title         = {{Back to the features: assessing the discriminating power of future CMB missions on inflationary models}},
  volume        = {03},
  year          = {2023}
}

@misc{cranmer2023,
  archiveprefix = {arXiv},
  author        = {Miles Cranmer},
  eprint        = {2305.01582},
  primaryclass  = {astro-ph.IM},
  title         = {Interpretable Machine Learning for Science with PySR and SymbolicRegression.jl},
  url           = {https://arxiv.org/abs/2305.01582},
  year          = {2023}
}

@article{Easther:2004vq,
  archiveprefix = {arXiv},
  author        = {Easther, Richard and Kinney, William H and Peiris, Hiranya},
  doi           = {10.1088/1475-7516/2005/05/009},
  eprint        = {astro-ph/0412613},
  journal       = {JCAP},
  pages         = {009},
  title         = {{Observing trans-Planckian signatures in the cosmic microwave background}},
  volume        = {05},
  year          = {2005}
}

@article{Easther:2013kla,
  archiveprefix = {arXiv},
  author        = {Easther, Richard and Flauger, Raphael},
  doi           = {10.1088/1475-7516/2014/02/037},
  eprint        = {1308.3736},
  journal       = {JCAP},
  pages         = {037},
  primaryclass  = {astro-ph.CO},
  title         = {{Planck Constraints on Monodromy Inflation}},
  volume        = {02},
  year          = {2014}
}

@article{Gong:2005jr,
  archiveprefix = {arXiv},
  author        = {Gong, Jinn-Ouk},
  doi           = {10.1088/1475-7516/2005/07/015},
  eprint        = {astro-ph/0504383},
  journal       = {JCAP},
  pages         = {015},
  reportnumber  = {KAIST-TH-2005-06},
  title         = {{Breaking scale invariance from a singular inflaton potential}},
  volume        = {07},
  year          = {2005}
}

@article{Guth:1980zm,
  author       = {Guth, Alan H.},
  doi          = {10.1103/PhysRevD.23.347},
  editor       = {Fang, Li-Zhi and Ruffini, R.},
  journal      = {Phys. Rev. D},
  pages        = {347--356},
  reportnumber = {SLAC-PUB-2576},
  title        = {{The Inflationary Universe: A Possible Solution to the Horizon and Flatness Problems}},
  volume       = {23},
  year         = {1981}
}

@article{Linde:1981mu,
  author       = {Linde, Andrei D.},
  doi          = {10.1016/0370-2693(82)91219-9},
  editor       = {Fang, Li-Zhi and Ruffini, R.},
  journal      = {Phys. Lett. B},
  pages        = {389--393},
  reportnumber = {LEBEDEV-81-229},
  title        = {{A New Inflationary Universe Scenario: A Possible Solution of the Horizon, Flatness, Homogeneity, Isotropy and Primordial Monopole Problems}},
  volume       = {108},
  year         = {1982}
}

@article{Meerburg:2013cla,
  archiveprefix = {arXiv},
  author        = {Meerburg, P. Daniel and Spergel, David N. and Wandelt, Benjamin D.},
  doi           = {10.1103/PhysRevD.89.063536},
  eprint        = {1308.3704},
  journal       = {Phys. Rev. D},
  number        = {6},
  pages         = {063536},
  primaryclass  = {astro-ph.CO},
  title         = {{Searching for oscillations in the primordial power spectrum. I. Perturbative approach}},
  volume        = {89},
  year          = {2014}
}

@article{Meerburg:2013dla,
  archiveprefix = {arXiv},
  author        = {Meerburg, P. Daniel and Spergel, David N.},
  doi           = {10.1103/PhysRevD.89.063537},
  eprint        = {1308.3705},
  journal       = {Phys. Rev. D},
  number        = {6},
  pages         = {063537},
  primaryclass  = {astro-ph.CO},
  title         = {{Searching for oscillations in the primordial power spectrum. II. Constraints from Planck data}},
  volume        = {89},
  year          = {2014}
}

@article{Mergulhao:2023ukp,
  archiveprefix = {arXiv},
  author        = {Mergulh{\~a}o, Thiago and Beutler, Florian and Peacock, John A.},
  doi           = {10.1088/1475-7516/2023/08/012},
  eprint        = {2303.13946},
  journal       = {JCAP},
  pages         = {012},
  primaryclass  = {astro-ph.CO},
  title         = {{Primordial feature constraints from BOSS + eBOSS}},
  volume        = {08},
  year          = {2023}
}

@article{Peiris:2013opa,
  archiveprefix = {arXiv},
  author        = {Peiris, Hiranya and Easther, Richard and Flauger, Raphael},
  doi           = {10.1088/1475-7516/2013/09/018},
  eprint        = {1303.2616},
  journal       = {JCAP},
  pages         = {018},
  primaryclass  = {astro-ph.CO},
  reportnumber  = {NSF-KITP-13-083},
  title         = {{Constraining Monodromy Inflation}},
  volume        = {09},
  year          = {2013}
}

@article{Peng:2025vda,
  archiveprefix = {arXiv},
  author        = {Peng, Ze-Yu and Piao, Yun-Song},
  doi           = {10.1103/zdvg-1ykz},
  eprint        = {2507.17276},
  journal       = {Phys. Rev. D},
  number        = {2},
  pages         = {023544},
  primaryclass  = {astro-ph.CO},
  title         = {{Tightening constraints on primordial oscillations with latest ACT and SPT data}},
  volume        = {113},
  year          = {2026}
}

@article{Petretti:2024mjy,
  archiveprefix = {arXiv},
  author        = {Petretti, Catherine and Braglia, Matteo and Chen, Xingang and Kumar Hazra, Dhiraj and Paban, Sonia},
  doi           = {10.1088/1475-7516/2025/06/035},
  eprint        = {2411.03459},
  journal       = {JCAP},
  number        = {035},
  pages         = {035},
  primaryclass  = {astro-ph.CO},
  title         = {{Investigating the origin of CMB large-scale features using LiteBIRD and CMB-S4}},
  volume        = {06},
  year          = {2025}
}

@article{Planck:2018jri,
  archiveprefix = {arXiv},
  author        = {Akrami, Y. and others},
  collaboration = {Planck},
  doi           = {10.1051/0004-6361/201833887},
  eprint        = {1807.06211},
  journal       = {Astron. Astrophys.},
  pages         = {A10},
  primaryclass  = {astro-ph.CO},
  title         = {{Planck 2018 results. X. Constraints on inflation}},
  volume        = {641},
  year          = {2020}
}

@article{Planck:2018vyg,
  archiveprefix = {arXiv},
  author        = {Aghanim, N. and others},
  collaboration = {Planck},
  doi           = {10.1051/0004-6361/201833910},
  eprint        = {1807.06209},
  journal       = {Astron. Astrophys.},
  note          = {[Erratum: Astron.Astrophys. 652, C4 (2021)]},
  pages         = {A6},
  primaryclass  = {astro-ph.CO},
  title         = {{Planck 2018 results. VI. Cosmological parameters}},
  volume        = {641},
  year          = {2020}
}

@article{schmidt2009distilling,
  author    = {Schmidt, Michael and Lipson, Hod},
  doi       = {10.1126/science.1165893},
  journal   = {Science},
  number    = {5923},
  pages     = {81--85},
  publisher = {American Association for the Advancement of Science},
  title     = {Distilling free-form natural laws from experimental data},
  volume    = {324},
  year      = {2009}
}

@article{Slosar:2019gvt,
  archiveprefix = {arXiv},
  author        = {Slosar, Anze and others},
  eprint        = {1903.09883},
  journal       = {Bull. Am. Astron. Soc.},
  number        = {3},
  pages         = {98},
  primaryclass  = {astro-ph.CO},
  title         = {{Scratches from the Past: Inflationary Archaeology through Features in the Power Spectrum of Primordial Fluctuations}},
  volume        = {51},
  year          = {2019}
}

@article{SPT-3G:2025bzu,
  archiveprefix = {arXiv},
  author        = {Camphuis, E. and others},
  collaboration = {SPT-3G},
  doi           = {10.1103/7wt3-9v2y},
  eprint        = {2506.20707},
  journal       = {Phys. Rev. D},
  number        = {8},
  pages         = {083504},
  primaryclass  = {astro-ph.CO},
  reportnumber  = {FERMILAB-PUB-25-0144-PPD},
  title         = {{SPT-3G D1: CMB temperature and polarization power spectra and cosmology from 2019 and 2020 observations of the SPT-3G main field}},
  volume        = {113},
  year          = {2026}
}

@article{Starobinsky:1980te,
  author  = {Starobinsky, Alexei A.},
  doi     = {10.1016/0370-2693(80)90670-X},
  editor  = {Khalatnikov, I. M. and Mineev, V. P.},
  journal = {Phys. Lett. B},
  pages   = {99--102},
  title   = {{A New Type of Isotropic Cosmological Models Without Singularity}},
  volume  = {91},
  year    = {1980}
}

@article{Starobinsky:1992ts,
  author  = {Starobinsky, Alexei A.},
  journal = {JETP Lett.},
  pages   = {489--494},
  title   = {{Spectrum of adiabatic perturbations in the universe when there are singularities in the inflation potential}},
  volume  = {55},
  year    = {1992}
}

@article{Wang:2002hf,
  archiveprefix = {arXiv},
  author        = {Wang, Xiulian and Feng, Bo and Li, Mingzhe and Chen, Xue-Lei and Zhang, Xinmin},
  doi           = {10.1142/S0218271805006985},
  eprint        = {astro-ph/0209242},
  journal       = {Int. J. Mod. Phys. D},
  pages         = {1347},
  title         = {{Natural inflation, Planck scale physics and oscillating primordial spectrum}},
  volume        = {14},
  year          = {2005}
}

@article{Sousa-Neto:2025gpj,
  archiveprefix = {arXiv},
  author        = {Sousa-Neto, Agripino and Bengaly, Carlos and Gonzalez, Javier E. and Alcaniz, Jailson},
  doi           = {10.1016/j.dark.2025.102108},
  eprint        = {2502.10506},
  journal       = {Phys. Dark Univ.},
  pages         = {102108},
  primaryclass  = {astro-ph.CO},
  title         = {{Symbolic regression analysis of dynamical dark energy with DESI-DR2 and SN data}},
  volume        = {50},
  year          = {2025}
}

@article{Koksbang:2026wvh,
  archiveprefix = {arXiv},
  author        = {Koksbang, S. M. and Heinesen, A.},
  eprint        = {2604.05822},
  month         = {4},
  primaryclass  = {astro-ph.CO},
  title         = {{Model-independent constraints on generalized FLRW consistency relations with bootstrap-based symbolic regression}},
  year          = {2026}
}

@article{Bartlett:2022kyi,
  archiveprefix = {arXiv},
  author        = {Bartlett, Deaglan J. and Desmond, Harry and Ferreira, Pedro G.},
  doi           = {10.1109/TEVC.2023.3280250},
  eprint        = {2211.11461},
  journal       = {IEEE Trans. Evol. Comput.},
  number        = {4},
  pages         = {964},
  primaryclass  = {astro-ph.CO},
  title         = {{Exhaustive Symbolic Regression}},
  volume        = {28},
  year          = {2024}
}

@article{Stahl:2025qru,
  archiveprefix = {arXiv},
  author        = {Stahl, Cl{\'e}ment and Werth, Denis and Poulin, Vivian},
  doi           = {10.1103/PhysRevD.111.123514},
  eprint        = {2502.02571},
  journal       = {Phys. Rev. D},
  number        = {12},
  pages         = {123514},
  primaryclass  = {astro-ph.CO},
  title         = {{Primordial sharp features through the nonlinear regime of structure formation}},
  volume        = {111},
  year          = {2025}
}

@article{Calderon:2025xod,
  archiveprefix = {arXiv},
  author        = {Calderon, Rodrigo and Simon, Th{\'e}o and Shafieloo, Arman and Hazra, Dhiraj Kumar},
  doi           = {10.1088/1475-7516/2026/01/057},
  eprint        = {2504.06183},
  journal       = {JCAP},
  pages         = {057},
  primaryclass  = {astro-ph.CO},
  title         = {{Primordial features in light of the Effective Field Theory of Large-Scale Structure}},
  volume        = {01},
  year          = {2026}
}

@article{Blas:2011rf,
  archiveprefix = {arXiv},
  author        = {Blas, Diego and Lesgourgues, Julien and Tram, Thomas},
  doi           = {10.1088/1475-7516/2011/07/034},
  eprint        = {1104.2933},
  journal       = {JCAP},
  pages         = {034},
  primaryclass  = {astro-ph.CO},
  reportnumber  = {CERN-PH-TH-2011-082, LAPTH-010-11},
  title         = {{The Cosmic Linear Anisotropy Solving System (CLASS) II: Approximation schemes}},
  volume        = {07},
  year          = {2011}
}

@article{Torrado:2020dgo,
  archiveprefix = {arXiv},
  author        = {Torrado, Jesus and Lewis, Antony},
  doi           = {10.1088/1475-7516/2021/05/057},
  eprint        = {2005.05290},
  journal       = {JCAP},
  pages         = {057},
  primaryclass  = {astro-ph.IM},
  reportnumber  = {TTK-20-15},
  title         = {{Cobaya: Code for Bayesian Analysis of hierarchical physical models}},
  volume        = {05},
  year          = {2021}
}

@article{Chen:2012ja,
  archiveprefix = {arXiv},
  author        = {Chen, Xingang and Ringeval, Christophe},
  doi           = {10.1088/1475-7516/2012/08/014},
  eprint        = {1205.6085},
  journal       = {JCAP},
  pages         = {014},
  primaryclass  = {astro-ph.CO},
  title         = {{Searching for Standard Clocks in the Primordial Universe}},
  volume        = {08},
  year          = {2012}
}

@article{Piao:2004tq,
  archiveprefix = {arXiv},
  author        = {Piao, Yun-Song and Zhang, Yuan-Zhong},
  doi           = {10.1103/PhysRevD.70.063513},
  eprint        = {astro-ph/0401231},
  journal       = {Phys. Rev. D},
  pages         = {063513},
  title         = {{Phantom inflation and primordial perturbation spectrum}},
  volume        = {70},
  year          = {2004}
}

@article{Cai:2020qpu,
  archiveprefix = {arXiv},
  author        = {Cai, Yong and Piao, Yun-Song},
  doi           = {10.1103/PhysRevD.103.083521},
  eprint        = {2012.11304},
  journal       = {Phys. Rev. D},
  number        = {8},
  pages         = {083521},
  primaryclass  = {gr-qc},
  title         = {{Intermittent null energy condition violations during inflation and primordial gravitational waves}},
  volume        = {103},
  year          = {2021}
}

@article{Baldi:2005gk,
  archiveprefix = {arXiv},
  author        = {Baldi, Marco and Finelli, Fabio and Matarrese, Sabino},
  doi           = {10.1103/PhysRevD.72.083504},
  eprint        = {astro-ph/0505552},
  journal       = {Phys. Rev. D},
  pages         = {083504},
  title         = {{Inflation with violation of the null energy condition}},
  volume        = {72},
  year          = {2005}
}

@article{SimonsObservatory:2018koc,
  archiveprefix = {arXiv},
  author        = {Ade, Peter and others},
  collaboration = {Simons Observatory},
  doi           = {10.1088/1475-7516/2019/02/056},
  eprint        = {1808.07445},
  journal       = {JCAP},
  pages         = {056},
  primaryclass  = {astro-ph.CO},
  title         = {{The Simons Observatory: Science goals and forecasts}},
  volume        = {02},
  year          = {2019}
}

\newpage
\clearpage

\appendix
\onecolumngrid

\begin{center}
    {\bf Supplemental Material for ``Inverse-k Primordial Oscillations from a Symbolic Regression Search''}
\end{center}

\section{Details of the SR search}

We write the primordial power spectrum (PPS) as
\begin{equation}
    P_{\mathcal R}(k) =
    P_{\mathcal R,0}(k)\left[1+f(k)\right]
\end{equation}
and $f(k)$ is the primordial feature we search for.
We use $k$ in ${\mathrm{Mpc}^{-1}}$ as the input variable of \texttt{PySR}.
For a given expression $f(k)$, the primordial spectrum is supplied to CLASS as an external spectrum and the
objective function is the CMB $\chi^2$ with the background and nuisance parameters fixed to their best-fit values.

The operator set used in the SR search is
\begin{equation}
    +,\quad -,\quad \times,\quad /,\quad \text{\textasciicircum},
\end{equation}
and
\begin{equation}
    \exp,\quad \log,\quad \sin,\quad \cos,\quad \tanh .
\end{equation}
Nested trigonometric functions are forbidden, and the maximum expression
size is set to 25.

The finite sampling limits the frequency that can be reliably resolved.
For the sake of efficiency, we set \texttt{k\_per\_decade\_primordial=400} in CLASS.
For an inverse-$k$ phase, \(\phi(k)=B/k\),
the phase variation over one CLASS sampling step is approximately
\begin{equation}
    |\Delta\phi|
    \simeq \frac{B}{k}\Delta\ln k,
    \quad
    \Delta\ln k=\frac{\ln 10}{400}.
\end{equation}
For $B\simeq4\,{\mathrm{Mpc}^{-1}}$, this gives $|\Delta\phi|\lesssim1$ for
$k\gtrsim0.02\,{\mathrm{Mpc}^{-1}}$. The
inverse-$k$ C8 solutions are therefore numerically stable over the main
CMB-sensitive range. Higher-complexity Planck
expressions containing inverse powers such as $1/k^2$ oscillate faster
at small $k$ and are not considered as robust physical results.

\section{Full SR results}

\begin{table}[htbp]
    \centering
    \footnotesize
    \begin{tabular}{|c|c|l|}
        \hline
        Complexity & $\chi^2_{\mathrm{Planck}}$ & $f(k)$ \\
        \hline
        1  & 606.52 & $2.90\times10^{-4}$ \\
        5  & 606.40 & $-0.00957 k^2$ \\
        6  & 605.51 & $0.00197/\cos k$ \\
        7  & 604.13 & $\sin(1.110 k^2)$ \\
        8  & 596.84 & $-0.0299\cos(4.101/k)$ \\
        10 & 595.08 & $-0.0339\cos(4.150/k-0.859)$ \\
        11 & 593.42 & $k\sin(-0.868/k^2)$ \\
        13 & 591.82 & $k\sin[-0.868/(k^2+8.89\times10^{-6}k)]$ \\
        16 & 585.17 & $0.0924\sin[(2.990+0.5248/k)/k]$ \\
        18 & 585.17 & $0.0924\sin(2.990/k+0.5248/k^2)$ \\
        19 & 585.15 & $0.0924\sin[(0.5248+2.990k)/k^2]$ \\
        \hline
    \end{tabular}
    \caption{Full Pareto-front expressions from the Planck SR run.}
    \label{tab:app_planck}
\end{table}

\begin{table}[htbp]
    \centering
    \footnotesize
    \begin{tabular}{|c|c|l|}
        \hline
        Complexity & $\chi^2_{\mathrm{SPA}}$ & $f(k)$ \\
        \hline
        1  & 463.90 & $2.22\times10^{-4}$ \\
        3  & 463.84 & $0.00527 k$ \\
        5  & 463.79 & $0.0133 k-7.31\times10^{-4}$ \\
        7  & 457.85 & $-0.0106\sin(-390.0 k)$ \\
        8  & 451.70 & $-0.0209\sin(4.231/k)$ \\
        10 & 451.33 & $0.0209\sin(4.178/k-2.594)$ \\
        15 & 450.67 & $0.0218\sin(4.173/k+4.173/e^k)$ \\
        24 & 450.34 & $0.0219\sin(4.165/k+4.266/e^k)$ \\
        \hline
    \end{tabular}
    \caption{Full Pareto-front expressions from the SPA SR run.}
    \label{tab:app_spa}
\end{table}

The full Pareto-front expressions from the Planck and SPA SR runs are
listed in Tables~\ref{tab:app_planck} and \ref{tab:app_spa},
respectively.

\end{document}